\documentclass[iop,twocolumn,natbib209]{emulateapj}  
\usepackage[]{natbib, graphicx, hyperref, CJK}
\usepackage{ulem}

\shorttitle{SMA observations of C$_2$H in High-Mass Star Forming Regions}
\shortauthors{Jiang et al.}
\slugcomment{June 2015, ApJ accepted}
\begin{document}
\begin{CJK*}{UTF8}{gbsn}
\title{SMA observations of C$_2$H in High-Mass Star Forming Regions}

\author{
Xue-Jian Jiang(蒋雪健)\altaffilmark{1,2,3,4},
Hauyu Baobab Liu\altaffilmark{5},
Qizhou Zhang\altaffilmark{2},
Junzhi Wang\altaffilmark{6},
Zhi-Yu Zhang\altaffilmark{7,8},
Juan Li\altaffilmark{6},
Yu Gao \altaffilmark{1} and  
Qiusheng Gu\altaffilmark{3,4} \\
}

\altaffiltext{1}{Purple Mountain Observatory \& Key Laboratory for Radio Astronomy,
Chinese Academy of Sciences, 2 West Beijing Road, Nanjing 210008, P.R.China}
\altaffiltext{2}{Harvard-Smithsonian Center for Astrophysics, 60 Garden St., 
Cambridge, MA 02138, USA}
\altaffiltext{3}{Key Laboratory of Modern Astronomy and Astrophysics, Nanjing
University, Ministry of Education, Nanjing 210093, P.R.China}
\altaffiltext{4}{Collaborative Innovation Center of Modern Astronomy and Space
Exploration, Nanjing 210093, China}
\altaffiltext{5}{Academia Sinica Institute of Astronomy and Astrophysics, P.O.
Box 23-141, Taipei 106, Taiwan}
\altaffiltext{6}{Shanghai Astronomical Observatory, Chinese Academy of
 Sciences, 80 Nandan Road, Shanghai 200030, P.R.China}
\altaffiltext{7}{Institute for Astronomy, University of Edinburgh, Royal
Observatory, Blackford Hill, Edinburgh EH9 3HJ, United Kindom}
\altaffiltext{8}{ESO, Karl Schwarzschild Strasse 2, D-85748 Garching, Munich,
Germany}
\email{Email: xjjiang@pmo.ac.cn}

% Common terms
\def\Ms{$M_{\rm *}$}
\def\Lsun{$L_{\odot}$}
\def\Hi{H\,{\sc i}~}				% HI
\def\Ht{H$_2$\,}                    % H2
\def\COto{$^{12}$CO(J=2$\rightarrow$1)\ }
\def\cch{C$_2$H }
\def\c2h2{C$_2$H$_2$}
\def\hcccn{HC$_3$N~}

% Units
\def\kms{$\rm km\,s^{-1}$}
\def\micro{\,$\mu$m}
\def\Msun{$M_{\odot}$}

% luminosity: 
% G10, W33 ~ 10^6 solar luminosity
% ON1 ~ 10^4 solar luminosity 
% AFGL 490 ~ 10^3 solar luminosity

\begin{abstract}
  C$_{2}$H is a representative hydrocarbon that is abundant and
  ubiquitous in the
  interstellar medium (ISM). To study its chemical properties, we present
  Submillimeter Array (SMA) observations of the C$_{2}$H $N$=3-2 and HC$_{3}$N
  $J$=30-29 transitions and the 1.1 mm continuum emission toward four OB
  cluster-forming regions, AFGL\,490, ON\,1, W33\,Main, and G10.6-0.4, which
  cover a  bolometric luminosity range of $\sim$10$^{3}$--10$^{6}$
  $L_{\odot}$. We found that on large scales, the C$_{2}$H emission
  traces the dense molecular envelope. However, for all observed sources,
  the peaks of C$_{2}$H emission are offset by several times times 10$^{4}$
  AU from the peaks of
  1.1 mm continuum emission, where the most luminous stars are
  located. By comparing the distribution and profiles of C$_{2}$H hyperfine
  lines and the 1.1 mm continuum emission, we find that the C$_{2}$H column
  density (and abundance) around the 1.1\,mm continuum peaks
  is lower than those in the ambient gas envelope. Chemical models suggest
  that C$_{2}$H might be transformed to other species owing to increased
  temperature
  and density; thus, its reduced abundance could be
  the signpost of the heated molecular gas in the $\sim$10$^{4}$ AU vicinity
  around the embedded high-mass stars.  Our results support such theoretical
  prediction for centrally embedded $\sim$10$^{3}$--10$^{6}$ $L_{\odot}$ OB
  star-forming cores, while future higher-resolution observations are required
  to examine the C$_{2}$H transformation around the localized sites of
  high-mass star formation. 
\end{abstract}


% ------------------------------------------------------------------------------
\section{Introduction} 
\setcounter{footnote}{0}

% -----------------------------------------------------------------------------------------------------------------------------------
%{\bf The question of determining evolutionary stages }\\ 
Observations have shown that high-mass stars form in massive molecular clumps
on scales of 0.1--1 pc \citep[e.g.,][]{Palau:2007, Galvan-Madrid:2009,
Zhang:2009, Wang:2012, Liu:2011, Liu:2012a, Liu:2012b, Galvan-Madrid:2013,
Wang:2014, Lu:2014}.  The strong protostellar feedback in radiative pressure
and mechanical processes (e.g. outflows, stellar wind) starts to disturb the
ambient molecular gas while the massive stars are still deeply embedded
\citep{Zinnecker:2007}.  How the contracting clumps dynamically evolve under
the influence of the (proto)stellar feedback remains uncertain
\citep[e.g.][]{Beuther:2007, Krumholz:2014}.  Extensive modelings have
suggested that this feedback may leave chemical footprints (e.g.
\citealp{Lahuis:2000, Chapman:2009, Herbst:2009, Tan:2014}). High spatial
resolution observations of molecular lines from fast-evolving
(10$^{4}$--10$^{6}$ yr) clouds are therefore helpful to diagnose the history of
cloud contraction and massive star formation even in the embedded phases.

Ethynyl radical (C$_2$H) is the simplest hydrocarbon molecule with the
C$\equiv$C bond, which is closely related to the formation and transformation
of long carbon chains and other hydrocarbons \citep{Pilleri:2013}. C$_2$H and
its predecessor \c2h2 are crucial intermediates in the chemistry of aromatic
rings (e.g., C$_6$H$_6$) and polycyclic aromatic hydrocarbon (PAH,
\citealt{Tielens:2013}).  However, because of its symmetry, \c2h2 does not have
a permanent dipole moment, so \c2h2 lacks of rotational transitions, which
makes \c2h2 observationally challenging.  On the other hand, C$_2$H has many
strong emission lines accessible in millimeter bands.  In addition, the
hyperfine structure lines (HFS) of C$_2$H can provide estimates of optical
depth and magnetic field by measuring the Zeeman effect. Its dipole moment
\citep[0.77 Debye,][]{Wilson:1977} is about seven times higher than that of
CO, and it has a critical density of $\sim 10^5\,{\rm cm}^{-3}$ \citep{Sakai:2010b}.

C$_2$H was first detected in Galactic star-forming regions by
\citet{Tucker:1974}, who derived a comparable C$_2$H abundance with those of
HCN and HCO$^+$.  Later surveys found that C$_2$H is widespread over the inner
Galactic plane \citep{Liszt:1995}, and it mainly arises from relatively dense gas
(i.e. $n(\rm H_2)\sim 10^4-10^5~\rm cm^{-3}$, \citealt{Watt:1988}). C$_2$H has
been detected in all evolutionary stages of high-mass star formation, e.g.
infrared dark cloud (IRDC), prestellar cores, high-mass protostellar objects
(HMPOs) and hot molecular cores (HMCs) \citep{Wootten:1980, Beuther:2007,
Li:2012}. Moreover, extragalactic studies show that C$_2$H is also quite
strong, after those popular (dense) gas tracers HCN, HNC, HCO$^+$, CS, etc., so
it can be used in multi-molecule diagnosis on galactic environment
\citep[e.g.,][]{Meier:2005, Meier:2012, Meier:2015, Jiang:2011,
Imanishi:2014}.

Single-dish and interferometric observations toward HMPO and ultracompact
H\textsc{ii} (UC\,H\textsc{ii}) regions found that C$_2$H abundance is decreasing
toward the central hot cores \citep{Beuther:2008, Pilleri:2013}, which
suggests that C$_2$H can trace cold molecular gas in the early phase of star
formation (see also \citealt{Gerin:2011}).  Observations on prestellar cores
\citep{Padovani:2009} also found similar trends of C$_2$H abundance and
implied that the sizes of C$_2$H depletion holes are a few thousand AU.
\cite{Mookerjea:2012} compared the distributions and abundances of C$_2$H and
C$_3$ in DR\,21 (OH) and suggested the chemical (gas-phase) timescale for
these species to be $\sim 0.7-3$ Myr. The aforementioned results, however, are
still limited by spatial resolution and biased samples. A systematic survey
will help clarify the environmental effects and will help us understand the
formation and evolution of C$_2$H in massive star-forming regions. 

%{\bf brief introduction of our work in this paper}\\ 
In this paper, we present the Submillimeter Array (SMA)\footnote{Submillimeter
Array is a joint project between the Smithsonian Astrophysical Observatory and
the Academia Sinica Institute of Astronomy and Astrophysics and is funded by
the Smithsonian Institution and the Academia Sinica.} observations of the
C$_2$H $N=3-2$ transitions toward four massive star-forming regions, G10.6-0.4,
W33 Main, ON\,1, and AFGL\,490.   These samples were selected from the previous
single-dish mapping survey of C$_2$H 1-0 (87.317 GHz), HNC 1-0 (90.664 GHz),
and \hcccn 10-9 (90.979 GHz) in 27 high-mass star forming regions
\citep{Li:2012}, using the Purple Mountain Observatory 14m Telescope (PMO-14m;
HPBW$\sim$55\arcsec).  The accurate parallax distances have been measured
except for AFGL\,490, whose distance has an uncertainty of $\sim$50\%.  Details
of our observations and data reduction are given in Section \ref{sec:obs}.  We
present the observational results in Sections \ref{sec:maps}.  The analysis of
the C$_2$H optical depth and abundance profiles are provided in Section
\ref{sec:tau} and \ref{sec:abundance}.  The implications of our results are
discussed in Section \ref{sec:chemistry}.

% ---------- end of Introduction ------------------------------------
% -------------------------------------------------------------------

\begin{deluxetable*}{rrrccc}
\tablecolumns{6}
\tablewidth{0pc}
%\tabletypesize{\scriptsize}
\tablecaption{General observational properties of the sample}
\tablehead{
\colhead{(1)} & \colhead{(2)}  & \colhead{(3) } & \colhead{(4) } 
& \colhead{(5)}  & \colhead{(6)}\\
\colhead{Source} &
\multicolumn{2}{c}{R.A.  (J2000)   Dec} &
\colhead{Distance} & 
\colhead{Luminosity} &
\colhead{Int.time} 	\\
\colhead{} &
\colhead{(h\phn m\phn s)~ ~ ~} &
\colhead{\phn(\degr~\phn \arcmin ~\phn \arcsec)\phn}& 
\colhead{(kpc)}   & 
\colhead{($L_{\odot}$)} &
\colhead{(minutes)}  
}
%\hline
\startdata
G10.6-0.4 & 18:10:28.70 & $-$19:55:48.7 & 4.95 & 7$\times$10$^5$	& 120 \\
W33 Main  & 18:14:13.67 & $-$17:55:25.2 & 2.40 & 5.4$\times$10$^5$ & 107 \\
ON\,1       & 20:10:09.14 &    31:31:37.4 & 2.35 & 2.7$\times$10$^4$ & 120 \\
AFGL\,490 & 03:27:38.80 &    58:47:00.0 & 1.0 					& 2$\times10^3$	    & 100  
%\hline
\enddata
%\end{tabular}
\tablecomments{(1) Source names; (2) and (3) coordinates of the phase
centers; (4) Distances of the
sources. Parallax distances of G10.6-0.4 from \citet{Sanna:2013}, W33
from \citet{Immer:2013}, and ON1 from \citet{Xu:2013}; (5) Bolometric
luminosities, converted from previous values using the
latest parallax distances, if available; (6) On-source integration time.
}\label{tab:sample}
%\end{center}
\end{deluxetable*}

% -----------------------------------------------------------------------------------------------------------------------------------
\begin{deluxetable*}{rcccccc}
\tablecolumns{7}
\tablewidth{0pc}
%\tabletypesize{\scriptsize}
\tablecaption{Measurements}
\tablehead{
\colhead{(1)} & \colhead{(2)}  & \colhead{(3) } & \colhead{(4)} & \colhead{(5)}  &
\colhead{(6)} & \colhead{(7)} \\ %& \colhead{(8) } & \colhead{(9)} & \colhead{(10)}\\
\colhead{Source} &
\colhead{Beam} &
\colhead{RMS$_{\rm cont}$} & 
\colhead{RMS$_{\rm C_2H}$}  & 
\colhead{RMS$_{\rm HC_3N}$} &
%\colhead{$S_{\rm 1.1mm}$}  &
\colhead{$S_{\rm C_2H}$} &
\colhead{$\Delta v$\tablenotemark{a}} \\
\colhead{} &
\colhead{(\arcsec)} &
\colhead{(Jy\,beam$^{-1}$)} & 
\multicolumn{2}{c}{(Jy\,beam$^{-1}$km\,s$^{-1}$)} &
\colhead{(Jy\,km\,s$^{-1}$)}     &
\colhead{(km\,s$^{-1}$)} 
}
%\hline
\startdata
G10.6-0.4 & 7.3$\times$6.0 & 0.20 	& 8.0 & 2.0  & 590  & [$-$10.7, 2.8]  \\
W33 Main  & 7.2$\times$5.9 & 0.12	& 3.4 & 2.6  & 212  & [27.0, 40.5]   \\
ON\,1     & 6.5$\times$5.6 & 0.04 	& 0.97& 1.3  & 75.4 & [7.4, 14.6]   \\
AFGL\,490 & 8.9$\times$5.0 & 0.02 	& 0.83& \nodata  & 41.0 & [$-17.1, -11.7$] 
%\hline
\enddata
%\end{tabular}
\tablecomments{
(1) Source names; 
(2) Synthesized beam (uniform-weighted) of the images;
(3) 1 $\sigma$ RMS of the continuum images; 
(4) 1 $\sigma$ RMS of the C$_2$H $N$=3-2 moment zero maps;
(5) 1 $\sigma$ RMS of the HC$_3$N $J$=30-29 moment zero maps; 
(6) Integrated flux of C$_2$H $N$=3-2; 
(7) Velocity ranges of the C$_2$H 3-2 (7/2--5/2, 4--3)
Main line. Note that the rms values listed here are higher than the expected 
thermal noise in the data since they are limited by the dynamic range in the images.}
\label{tab:data}
%\end{center}
\end{deluxetable*}

% -----------------------------------------------------------------------------------------------------------------------------------
\section{Observations and Data Reduction} \label{sec:obs} 
%\subsection{The Sample}
%\subsection{Observation and Data Reduction}
The SMA observations were carried out on 2012 August 4, with six available
antennae in the subcompact configuration. The phase referencing and pointing centers coincide with the water
maser positions in the selected targets (see \citealt{Li:2012}). The weather
condition was moderate ($\tau_{\mbox{\scriptsize{225 GHz}}} \sim$ 0.25),
yielding a typical system temperature $T_{\mbox{\scriptsize{sys}}}$ of
$\sim$300 K. The passband calibrator is 3C 279, and the flux calibrators are
Titan/Callisto.  The absolute fluxes are accurate to within 15\%.

The observations covered a frequency range of about 259--263 GHz in the lower
sideband and 271--275 GHz in the upper sideband. Our observations covered the
C$_2$H $N$=3$\rightarrow$2, $J$=7/2$\rightarrow$5/2, $F$=4$\rightarrow$3,
$\nu_{\rm rest}$=262.00426 GHz (blended with the $F$=3$\rightarrow$2 line), and
$J$=5/2$\rightarrow$3/2, $F$=3$\rightarrow$2, $\nu_{\rm rest}$=262.06499 GHz
(blended with the $F$=4$\rightarrow$3 line). These are the two main groups of
the C$_2$H $N=3-2$ hyperfine structures.  We simultaneously obtained \hcccn
$J=30-29$ ($E_{u} \sim $ 203 K), which is an excellent tracer of HMCs
\citep{Chapman:2009, Meier:2012}.  A uniform spectral channel spacing
was set to be 0.8125 MHz ($\sim$ 0.93 km\,s$^{-1}$) across the entire passband.
The SMA data were calibrated using the \texttt{MIR} IDL software
package.\footnote{The MIR Cookbook by Chunhua Qi can be found at
\url{http://cfa-www.harvard.edu/~cqi/mircook.html}} We use the line-free
channels in both the lower and upper sidebands to generate the continuum
channels, which are then imaged jointly to produce the 1.1 mm continuum image.
The continuum emission is subtracted before performing the spectral line
imaging.  The imaging is carried out using the {\tt CASA} \citep{McMullin:2007}
package, with uniform weighting and hogbom CLEANing algorithm. We do not
perform primary beam corrections since the angular sizes of the emission are
small compared with the SMA primary beam ($\sim$1$'$). In addition, the
primary beam effects are factored out when we analyze the intensity ratios (see
section \ref{sec:abundance}). Table \ref{tab:sample} lists the observational
properties of these sources, including the integration time, and the bolometric
luminosities of these sources.  
% -----------------------------------------------------------------------------------------------------------------------------------

\begin{figure*} \begin{center}
  \includegraphics[angle=0,scale=0.74]{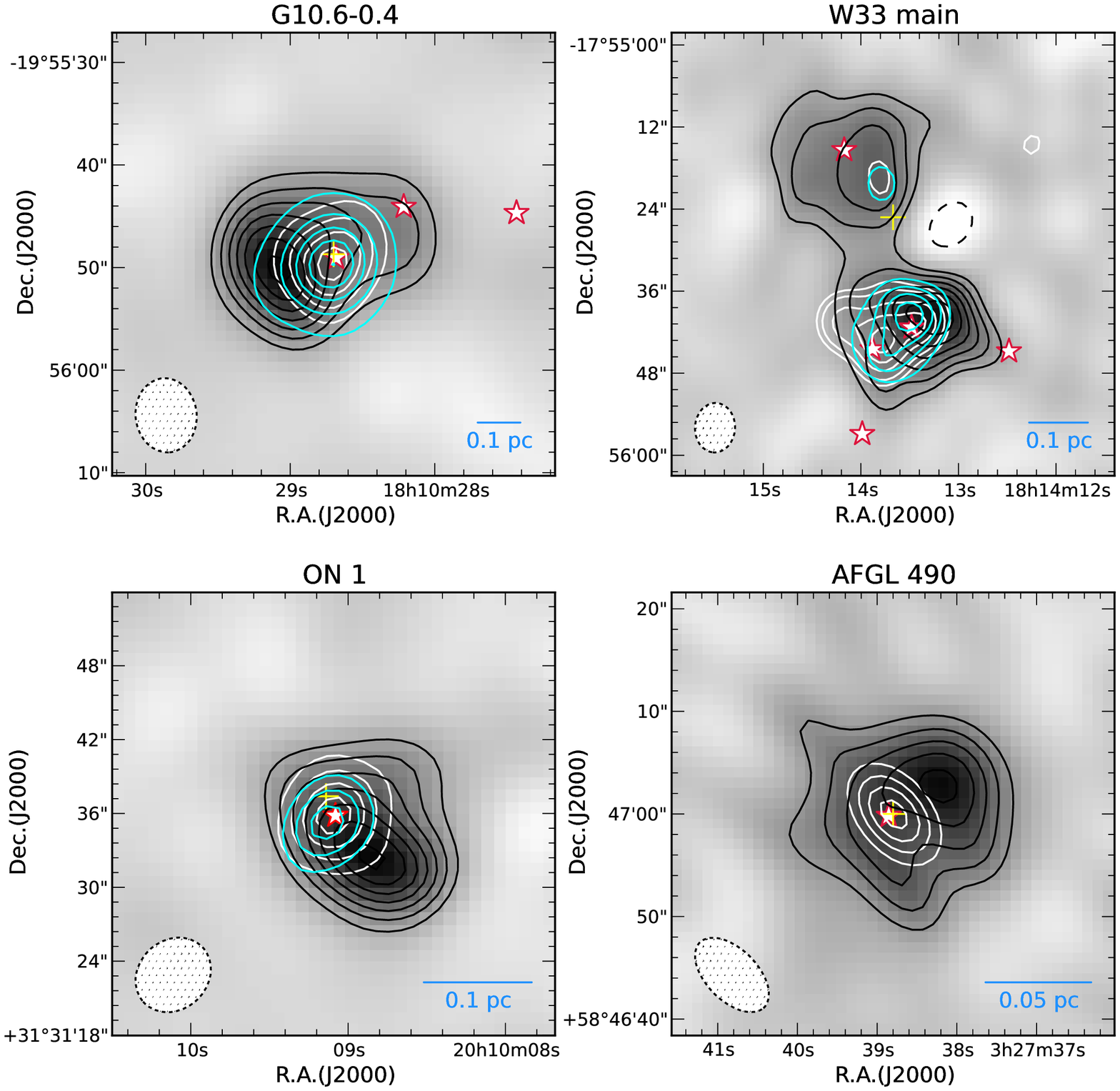} 
  \caption{Grayscale and
  black contours are the C$_2$H $N=3-2$ ($J=$7/2-5/2, $F=4-3$) uniformly weighted
  moment zero (integrated intensity) maps of G10.6-0.4, W33, ON\,1 and
  AFGL\,490, respectively.  C$_2$H intensity levels (black contours) start from
  the 5\,$\sigma$ level and continue in 3\,$\sigma$ steps.  The white contours are
  the 1.1 mm continuum images (30\%, 50\%, 70\%, and 90\% the peak emission). 
  The continuum peaks are 8.0, 4.1, 1.8, and 2.0 Jy beam$^{-1}$, respectively.
  The cyan contours are \hcccn $J=30-29$ uniformly weighted moment zero maps.
  In G10.6-0.4 the \hcccn 30-29 contours start from the 5\,$\sigma$ level and
  continue in 10\,$\sigma$ steps. In W33\,Main and ON\,1 the \hcccn 30-29
  contours start from the 4\,$\sigma$ level and continue in 3\,$\sigma$ steps.
  The
  1\,$\sigma$ of the continuum, C$_2$H 3--2 and \hcccn 30--29 are listed in Table
  \ref{tab:data}.  The red asterisks denote the positions of the
  (UC)\,H\,\textsc{ii} regions  \citep{Ho:1986, Haschick:1983, Kumar:2004,
  Simon:1983}.
  The yellow crosses denote the position of the
  H$_2$O masers.  The SMA subcompact synthesis beams of the C$_2$H images are
  shown in the bottom left corners. 
  \textit{A colorful figure is available in the online version.}}
  \label{fig:mom0} 
\end{center} 
\end{figure*}

\section{Results} \label{sec:results}
% -----------------------------------------------------------------------------------------------------------------------------------

\subsection{The C$_2$H N=3-2 brightness distribution}\label{sec:maps}

Figure \ref{fig:mom0} shows the integrated intensity maps of C$_{2}$H $N$=3-2
($E_u \sim$ 25 K), which are overlaid with the 1.1\,mm continuum emission
(white contours) and \hcccn $J$=30-29 integrated intensity (cyan contours).
The 1.1\,mm continuum emission may be  contaminated by both the dust
thermal emission and the free-free emission from the UC H\textsc{ii} regions.
The high-excitation \hcccn $J$=30-29 line ($E_{u} \sim $ 203 K) helps 
locate the HMCs. We resolved the extended C$_{2}$H emission surrounding
the continuum peaks (Figure \ref{fig:mom0}), while the previous single-dish
observations of the lower excited C$_2$H $N$=1-0 line ($E_u \sim$ 4 K)
traced the diffuse emission from a more extended area \citep{Li:2012}. The
size and geometry of the warm molecular gas traced by C$_{2}$H $N=3-2$ are
consistent with those traced by other molecular lines (e.g. G10.6-0.4:
\citealt{Liu:2010a}; W33\,Main:
\citealt{Haschick:1983}), suggesting that the resolved C$_2$H $N=3-2$
brightness distributions are likely not strongly biased by missing flux (Section
\ref{sec:chemistry}). The peaks of C$_2$H $N$=3-2 emission and continuum have
systematic offsets. We do not resolve ring-like C$_2$H distribution found in
\citet{Beuther:2008} and \citet{Guelin:1999}, likely because of the 
projection (or inclination) of the sample and the insufficient angular
resolution of the observations. 

\begin{figure*}[ht] \begin{center}
  \includegraphics[angle=0,scale=0.75]{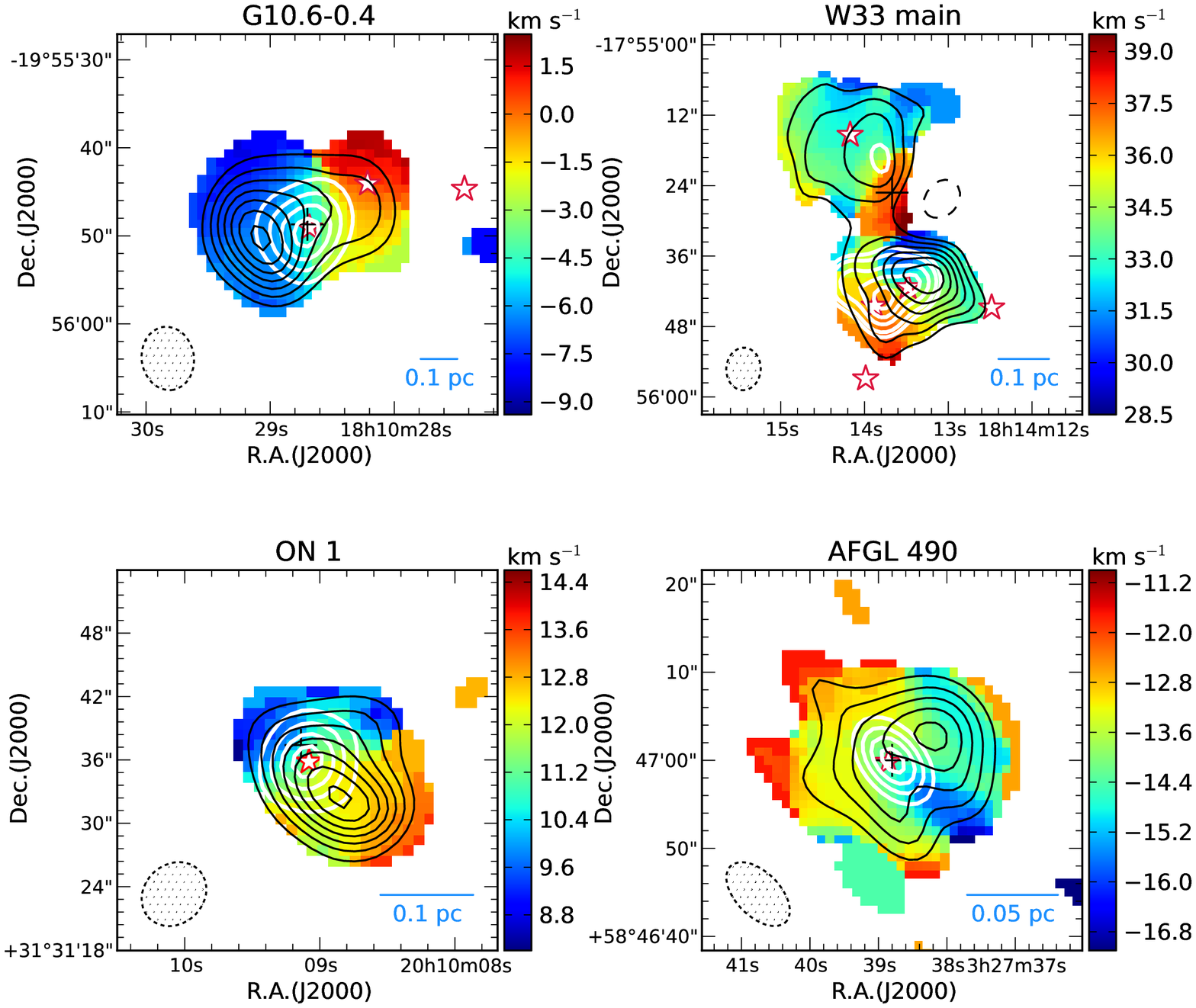} 
  \caption{Moment one (intensity-weighted average velocity) maps of C$_2$H
  $N=3-2$ ($J=$7/2--5/2, $F$=4--3) of G10.6-0.4, W33, ON\,1 and AFGL\,490. The
  black
  contours are the C$_2$H 3--2 moment zero, and the white contours are the
  1.1 mm continuum, as shown in Figure \ref{fig:mom0}.
  The asterisks and crosses denote the (UC)\,H\,\textsc{ii} regions and the
  H$_2$O masers.
  The SMA subcompact synthesis beams ($\sim 6\arcsec$) are shown in the bottom left corners.
  \textit{A colorful figure is available in the online version.}}
  \label{fig:mom1} 
\end{center} 
\end{figure*}

Figure \ref{fig:mom1} shows the intensity-weighted velocity maps.  The C$_2$H
3$-$2 emission shows velocity gradients, while its centroid velocities are
comparable with the systemic velocities.  More information about the individual
targets is provided as follows. The measurements of the rms noise of the maps
and the fluxes of C$_2$H 3-2 are listed in Table \ref{tab:data}. We achieve
$\sim$6$''$ angular resolution in the final images.

\begin{figure}[ht] \begin{center}
  \includegraphics[angle=0,scale=0.7]{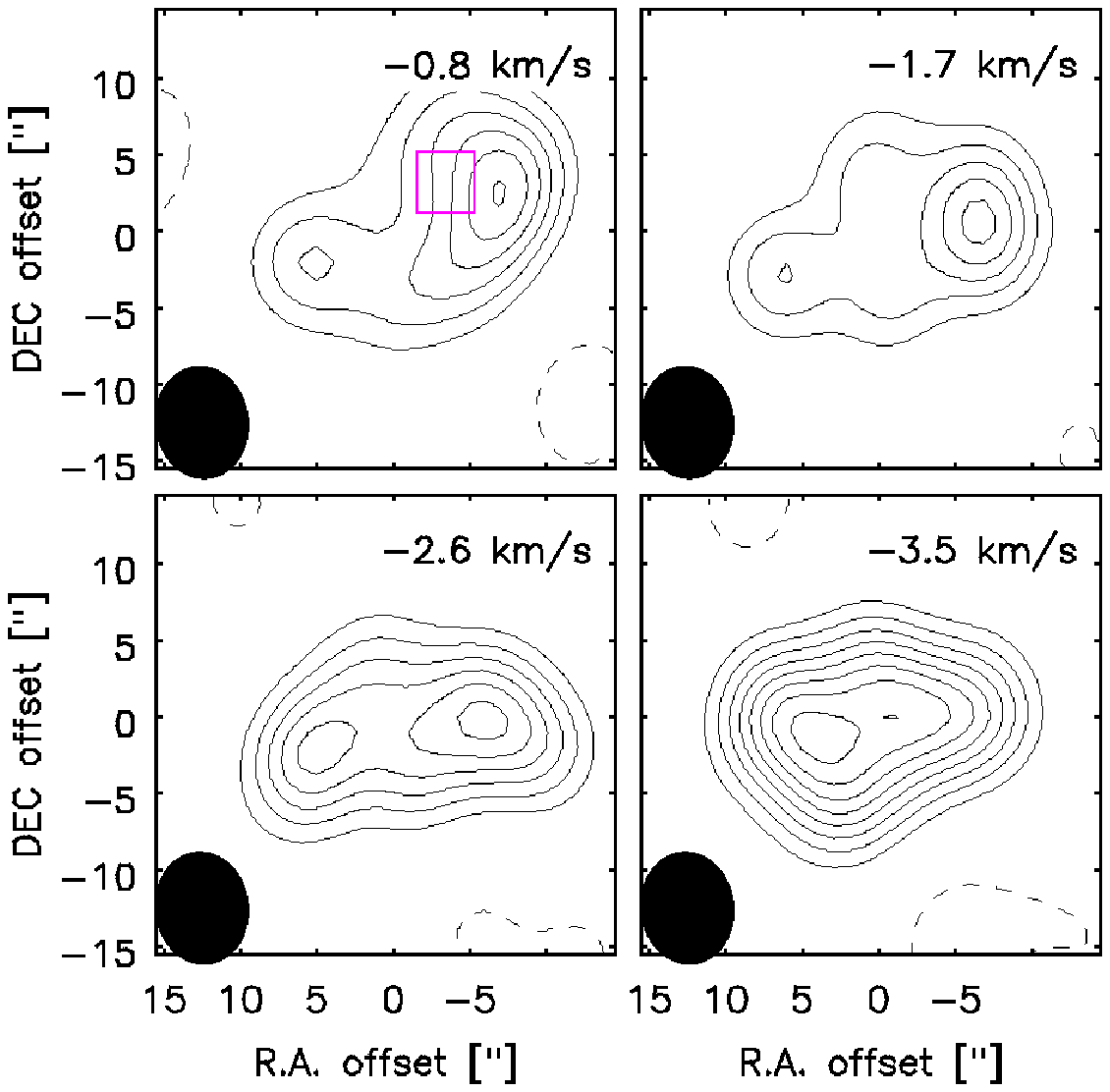}\\
  \includegraphics[angle=0,scale=0.38]{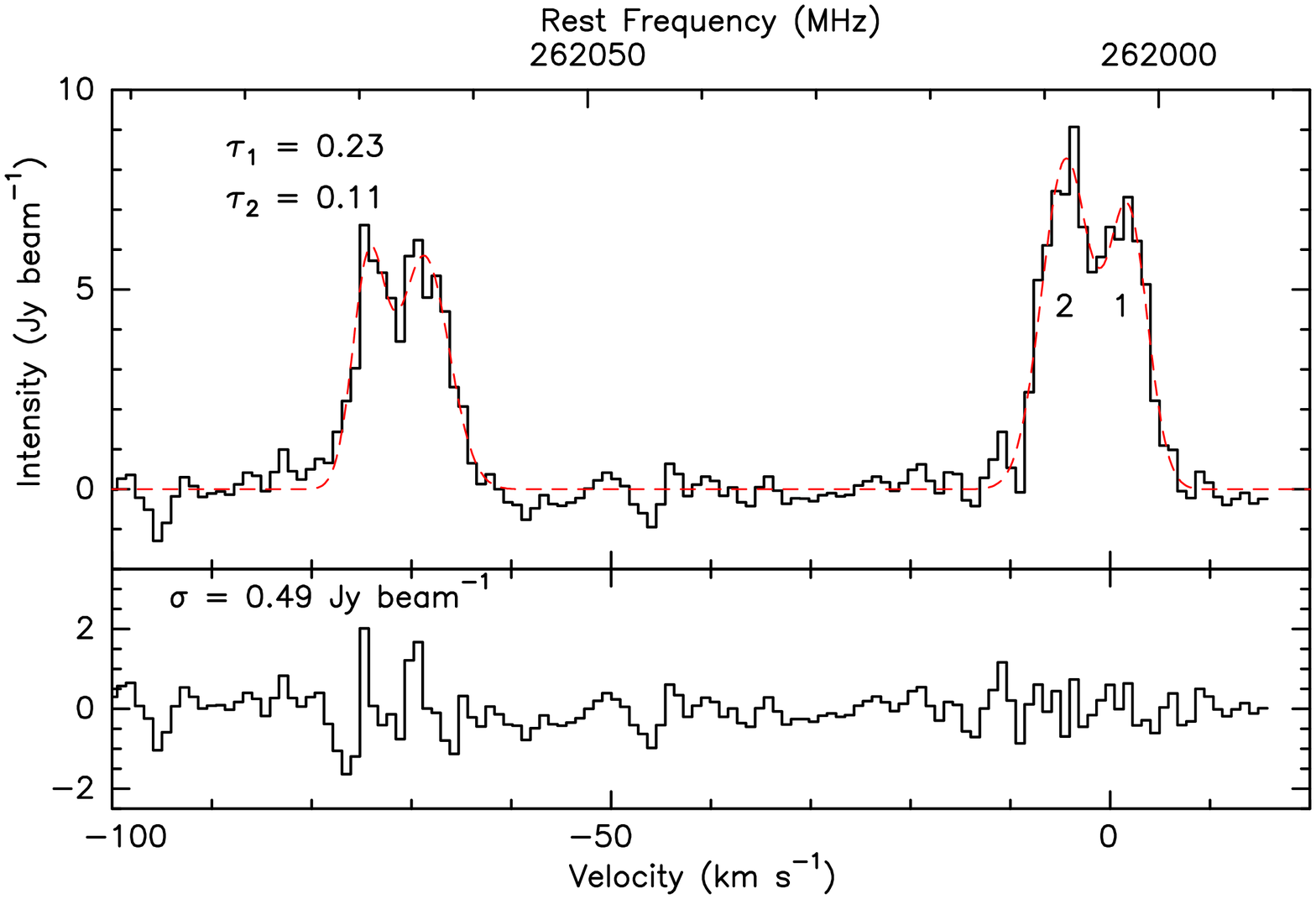} 
  \caption{Examples of C$_2$H $N=3-2$ channel maps of G10.6-0.4, from
  $V_{\rm LSR}$ = -0.8 km\,s$^{-1}$~to -3.5 km\,s$^{-1}$. Contours start from
  5\,$\sigma$ and increase in 3\,$\sigma$ steps (dashed contours are negative).
  The SMA subcompact synthesis beams are shown
  in the bottom left corners. A solid square in the first channel map denotes
  the position where we extract the spectrum in the bottom panel. The
  spectrum shows two velocity components of C$_2$H 3--2 emission in that
  position. Numbers ``1'' and ``2'' denote the main hyperfine groups of the
  two velocity components, and the red dashed line is the fitting to the
  spectrum. The fitting residual and corresponding RMS are shown below the
  spectrum. \textit{A colorful figure is available in the online version.}}
  \label{fig:channel} 
\end{center} 
\end{figure}

%------------------ G10.6 -------------------------
\paragraph{G10.6-0.4}  The UC\,H\textsc{ii} region G10.6-0.4 \citep{Ho:1986,
Keto:1987, Keto:1988, Sollins:2005, Liu:2010a, Liu:2010b, Liu:2011} is a
well-studied OB cluster-forming region \citep{Ho:1981, Liu:2013}, which is
deeply embedded in the convergence of several $\sim$5 pc scale dense gas
filaments. Adopting the parallax distance of 4.95 kpc \citep{Sanna:2013}, its
bolometric luminosity is about $ 7\times 10^5$ $L_{\odot}$ (note that the
luminosities for all the sources in this paper are recalculated with their
parallax distances, if available).

On the $\sim$0.5 pc scale, the C$_{2}$H $N=3-2$ emission (Figure
\ref{fig:mom1})  shows an obvious velocity gradient, and it likely traces the
edge-on spinning-up molecular accretion flow resolved by the previous
observations of highly excited NH$_{3}$ lines \citep{Ho:1986, Keto:1987,
Liu:2010a}.  The channel maps of C$_2$H $N=3-2$ (Figure \ref{fig:channel})
demonstrate the known southeast--northwest velocity gradient in this flattened
accretion flow \citep{Keto:1987, Liu:2010a}. The integrated intensity of
C$_{2}$H peaks at both blueshifted and redshifted lobes, which locate to the SE
and NW. It shows weaker emission at the center. We also notice that the C$_2$H
emission is weaker in the west of the region, where the other two
UC\,H\,\textsc{ii} regions reside (red asterisks in Figure \ref{fig:mom0} and
\ref{fig:mom1}).

%----------------- W33 Main -------------------------
\paragraph{W33\,Main} W33\,Main (G12.8-0.2) resides in the OB cluster-forming
region W33 Complex \citep{Haschick:1983}. At a distance of 2.4 kpc
\citep{Immer:2013}, its bolometric luminosity is about
$5.4\times10^5$\,$L_{\odot}$.  In our observations, the 1.1\,mm
continuum ($\sim$3 Jy) peaks at R.A.(2000) = {$\rm 18^h14^m13.8^s$},
decl.(2000) = $-17\degr55\arcmin43.4\arcsec$, which is associated with the
infrared source IRS\,3 \citep{Dyck:1977} and the centimeter peak W33 Main-B2.
The distribution of \hcccn 30--29, however, is slightly different and is closer to
another centimeter emission peak, B1 \citep{Haschick:1983}. We also detect another
1.1\,mm source associated with the IRS\,2 and the UC\,H\textsc{ii} region W33
Main-A in the north. It shows weaker emission in 1.1 mm continuum ($\sim$ 0.6 Jy)
and in the \hcccn 30--29 transition, comparing to the south component.

As shown in Figure \ref{fig:mom0}, the C$_{2}$H $N=3-2$ emission is mainly
distributed around W33\,Main-A and W33\,Main-B. The emission of C$_{2}$H 3-2,
\hcccn 30--29 and 1.1\,mm continuum around W33\,Main-B are brighter and appear
to be offset from each other, while in W33\,Main-A their peak positions agree
better with each other.  The environment of W33\,Main is quite complicated
since there are multiple (UC)H\,\textsc{ii} regions, and these
(UC)H\,\textsc{ii} regions A, B1, and B2 seem to be embedded in the gas envelope
traced by C$_{2}$H 3-2. Although the projection effect cannot be ruled out, the
differences between these two regions (W33\,Main-A and B) could be explained by
their different physical conditions and star-forming activities. Comparing with
W33\,Main-A, W33\,Main-B is much more massive and brighter in infrared,
millimeter, and centimeter continuum, as well as consisting of more H\textsc{ii}
regions \citep{Haschick:1983}. As revealed by the continuum and \hcccn on $\sim
0.1$ pc scales in Figure \ref{fig:mom0}, the different behaviors of B1 and B2
also suggest that these two regions have different stages of SF process. The
velocity field in W33\,Main-B is complicated (Figure \ref{fig:mom1}), which
suggests influence from the multiple molecular outflows \citep{Immer:2014}.

%----------------- ON 1 -------------------------
\paragraph{ON\,1} The UC\,H\textsc{ii} region Onsala\,1 (ON\,1) is a
cluster-forming region with multiple star-forming signposts \citep{Kumar:2003,
Su:2009}.
At a parallax distance of 2.35 kpc \citep{Xu:2013} its luminosity is about
$2.7\times 10^4$ $L_{\odot}$ \citep{MacLeod:1998}.  Figure \ref{fig:mom0} shows
that the 1.1 mm continuum and the \hcccn $J=30-29$ emission both agree well
with the center of the UC\,H\textsc{ii} and the H$_2$O maser, while the C$_2$H
$N$=3-2 emission peaks show offsets in the southwest. Figure \ref{fig:mom1}
shows that C$_2$H has a velocity gradient along the northeast--southwest
direction, and the velocity at the emission peak is consistent with the
systemic velocity $v_{\rm LSR}$=12 \kms~\citep{Su:2009}. Such distribution
along the NE--SW direction is in agreement with the H$^{13}$CO$^+$ distribution
reported by \citet{Kumar:2004}, which is interpreted as one of the multiple
molecular outflows in ON\,1, while high-resolution observations toward the
innermost region by \citet{Su:2009} favor a scenario of the expansion of the
H\textsc{ii} region.  Other than the main SW component, Figures \ref{fig:mom0}
and \ref{fig:mom1} also show that the C$_2$H emission appears to slightly
extend to the northwest, in a direction perpendicular to the proposed bipolar
outflow (or expansion motion).  This component might be related to the other
outflow component traced by SiO and CO as reported by \citet{Kumar:2004}. These
authors derived that the central enclosed mass of ON\,1 is about 300--400
\Msun, which would be sufficient to support a rotating structure, so with
current data we are not yet able to rule out the possibility of a flattened
rotating accretion flow confining the UC\,H\textsc{ii} region. 

%----------------- AFGL 490 -------------------------
\paragraph{AFGL\,490} AFGL\,490 is a deeply embedded 8--10 $M_{\odot}$ young
star.  Previous CO and $^{13}$CO observations suggest a warm ($T_{\rm kin}
\gtrsim $100 K) gas component within the central $\approx$ 6000 AU radius,
surrounded by a cooler gas envelope \citep{Mitchell:1995}.
\citet{Schreyer:2002} reported that the size of the outer envelope is
about 22,000 AU $\times$ 6000 AU.  Molecular outflows have also been inferred
from the detection of high-velocity CO emission \citep{Mitchell:1995}.

In AFGL\,490 C$_2$H 3--2 is significantly weaker than that in other sources,
and \hcccn 30-29 is not detected.  The brightness distribution of C$_{2}$H
$N=3-2$ mainly follows the flattened dense gas envelope traced by the CS
$J=3-2$ transition (Figure \ref{fig:mom0}; \citealt{Schreyer:2002}) and peaks
at northwest of the 1.1 mm continuum source. The velocity gradient of C$_{2}$H
$N$=3-2 in Figure \ref{fig:mom1} is not as clear as in G10.6-0.4 or ON\,1, but
it still shows that the C$_2$H $N$=3-2 emission consistently traces the
northwest--southeast velocity gradient in this flattened dense gas envelope, as
well as the molecular gas wind perpendicular to the flattened envelope
\citep{Schreyer:2002}.

\subsection{Optical depth of C$_2$H $N=3-2$}\label{sec:tau}
The electron--nucleus interaction of C$_2$H results in six splitting hyperfine
line components.  Optical depth can be obtained by fitting the hyperfine
structures.  We extract the C$_{2}$H $N=3-2$ spectra in every
$\sim$3$''$$\times$3$\arcsec$ pixels and fit the optical depth using the
\texttt{CLASS} package,\footnote {\url{http://www.iram.fr/IRAMFR/GILDAS}} based
on the theoretical intensity ratios provided in the CDMS
database\footnote{\url{http://www.astro.uni-koeln.de/cdms}}
\citep{Muller:2005}. LTE relative line intensities were assumed.  The bottom
panel of Figure \ref{fig:channel} shows an example spectrum of G10.6-0.4
extracted from a position offset from the core center (indicated by the solid
square in the first channel map of Fig.  \ref{fig:channel}). The spectrum shows
two velocity components (line separation $\Delta V \sim$ 5 km\,s$^{-1}$,
labeled `1' and `2'), which would cause confusion in the fitting of hyperfine
structures (the velocity separation of the two hyperfine lines of each group is
about 2.54 km\,s$^{-1}$).  We use Gaussian profiles to decompose the two
velocity components, and then we derive their optical depths separately. 

We found that C$_{2}$H is optically thin ($\tau \sim 0.1$) in most
regions, which is comparable with the optical depth of C$_2$H found in the
UC\,H\textsc{ii} region MonR2 \citep{Pilleri:2013}. Using a single-dish
telescope \citet{Li:2012} also found that the optical depths of C$_2$H are
small to moderate in most massive star-forming regions. The estimates of the
C$_{2}$H column density $N_{\rm C_2H}$ and abundance
$X_{\rm C_2H}$ are given in section \ref{sec:abundance}.

\begin{figure}[ht] 
  \begin{center}
  \includegraphics[angle=0,scale=0.43]{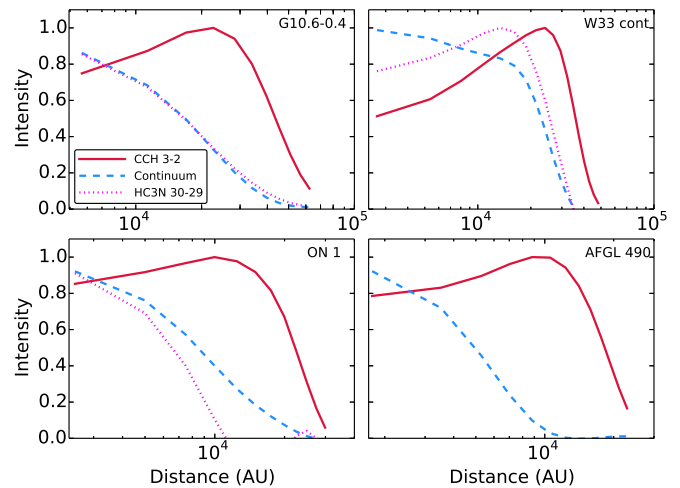} \caption{Normalized
  intensity profiles of different tracers ({\it red solid lines}: C$_2$H 3--2,
  {\it blue dashed lines}: continuum;  {\it
  magenta dotted lines}: \hcccn 30--29) of the four sources, which are extracted
  and measured along a strip starting from the continuum peaks to the C$_2$H
  peaks. Horizontal axis is the distance from the continuum peaks. 
  \textit{A colorful figure is available in the online version.}}
   \label{fig:profiles1}
\end{center} 
\end{figure}

\subsection{Intensity and column density profiles}\label{sec:abundance}
The \hcccn $J=30-29$ emission ($E_{u} \sim $ 203 K) provides a good probe of the
dense and warm gas concentration.  We did not detect \hcccn 30--29 in AFGL\,490.
For
the rest of the sources, the normalized 1.1 mm continuum and the radial
distributions of C$_2$H 3--2 and \hcccn 30--29 are presented in Figure
\ref{fig:profiles1}. We use one strip from the continuum peak to the C$_2$H
peak, to extract the intensity profile for each source, so as to show the
relative offset scale of different tracers.

The C$_2$H 3--2 intensity increases outward until several times 10$^4$
AU separation from the 1.1 mm emission peak and then drops steeply.  In
G10.6-0.4 and ON\,1, \hcccn 30-29 appears to be correlated with the 1.1 mm
continuum emission and is confined in the compact regions (see also Figure
\ref{fig:mom0}).  In W33\,Main, \hcccn 30--29 peaks in between the peaks of the
1.1 mm continuum and the C$_2$H 3-2 emission, and \hcccn is very close to one of
the H\,\textsc{ii} region B1 (Figure \ref{fig:mom0}). Nevertheless, the C$_2$H
3--2 emission peak is spatially offset from the 1.1\,mm continuum and
the \hcccn 30--29 emission peak in W33\,Main.

To quantify the column density and abundance profiles of C$_2$H, we extract the
C$_2$H 3-2 line intensity and the 1.1\,mm continuum flux in several regions
with diameter $\sim$ 3$\arcsec$ from the continuum peak and through the
C$_2$H 3-2 line emission
peak, until the S/N drops to $<2\,\sigma$ for either the 1.1\,mm continuum or
the C$_2$H 3-2 emission. Column densities of \Ht are derived from 1.1\,mm 
continuum, following
\citet{Beuther:2002,Beuther:2005} and assuming an emissivity $\beta = 1.5$
and $\kappa_{1.2 mm}$ = 0.97. 
Following \citet{Hunter:2006}, the column density of C$_2$H,
$N({\rm C_2H})$, is derived using 
\begin{equation} 
  N({\rm C_2H}) = \frac{8\pi k \nu^2}{h c^3 A_{ul}} \frac{Q_t}{g_u} \exp(\frac{E_u}{k 
  T_{ex}})\cdot S_{\rm C_2H}
\end{equation}
where $A_{ul}, Q_t, g_u$ are the Einstein $A$ coefficient, the partition function,
and the upper-state degeneracy quoted from the CDMS, respectively. $S_{\rm
C_2H}$ is the line intensity of C$_2$H in units of K\,km\,s$^{-1}$.
Identical $T_{\rm ex} = 20$ K and $T_{\rm dust} = 50$ K are adopted in the
calculations for all the sources.  Such an assumption on $T_{\rm dust}$ is
consistent with previous values adopted for these sources
\citep[e.g.,][]{Liu:2010a, Haschick:1983, Kumar:2004, Mueller:2002,
Schreyer:2006}, where the $T_{\rm ex}$ are found to be all close to 20\,K 
in the hyperfine structure fittings of C$_2$H (Section \ref{sec:tau}), similar
to the value derived by \citet{DeBeck:2012}. We find that the derived $N({\rm
C_2H})$ values are a few times $10^{15} {\rm cm^{-2}}$ for all targets.
 
There are several sources of errors in the estimates of the C$_2$H column
density and abundance relative to the \Ht molecule. The variation of $T_{\rm
ex}$ between 15 and 25 K can introduce up to 0.3 dex error in log $N({\rm
C_2H})$, so the typical error in log $N({\rm C_2H})$ is about 0.4 dex
considering the error on $S_{\rm C_2H}$ of about 50\%.  The variation of
$T_{\rm dust}$ between 40 and 60 K would also result in about 0.1 dex error in
log $N({\rm H_2})$. Although adopting different values of $\beta$ (e.g., $\beta
=$ 1, 1.5, 2) will introduce about 0.3 dex uncertainty in log $N({\rm H_2})$,
we expect that $\beta$ is probably similar in these sources.  Uncertainties on
the absolute fluxes of continuum and \cch line are factored out as $X({\rm
C_2H})$ is based on their intensity ratios. Thus, the typical error in log
$X({\rm C_2H})$ is about 0.5 dex. However, we caution that the derived
C$_{2}$H  column density and abundance profiles cannot be very precise at this
moment.  This is mainly because (1) part of the 1.1\,mm continuum may be
contaminated by free-free emission, which cannot be accurately subtracted yet,
and (2) the missing fluxes in interferometric observations can lead to uneven
underestimates of the continuum and line intensities. Thus, the results could be
possibly affected by another factor of two. Nevertheless, as we used the SMA
subcompact configuration with about 6$''$ resolution, those extended emissions
on a scale of less than 1$'$ \citep[$\sim 1$ pc;][]{Li:2012} should be
recovered effectively, so we suppose that the amount of diffuse emissions should be
limited and the results are not affected significantly. 

\begin{figure}[t] 
  \begin{center}
  \includegraphics[angle=0,scale=0.43]{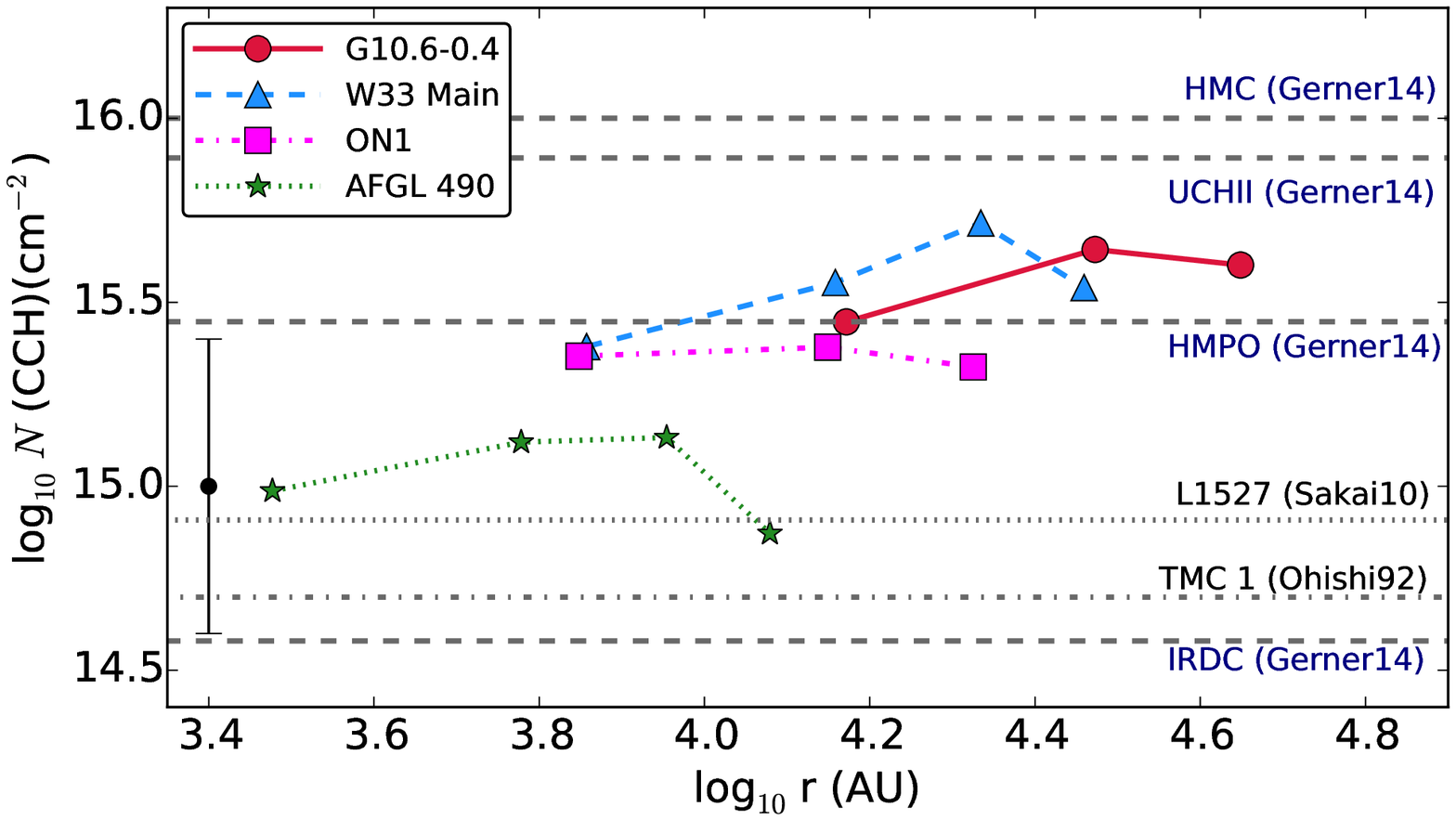} 
  \includegraphics[angle=0,scale=0.43]{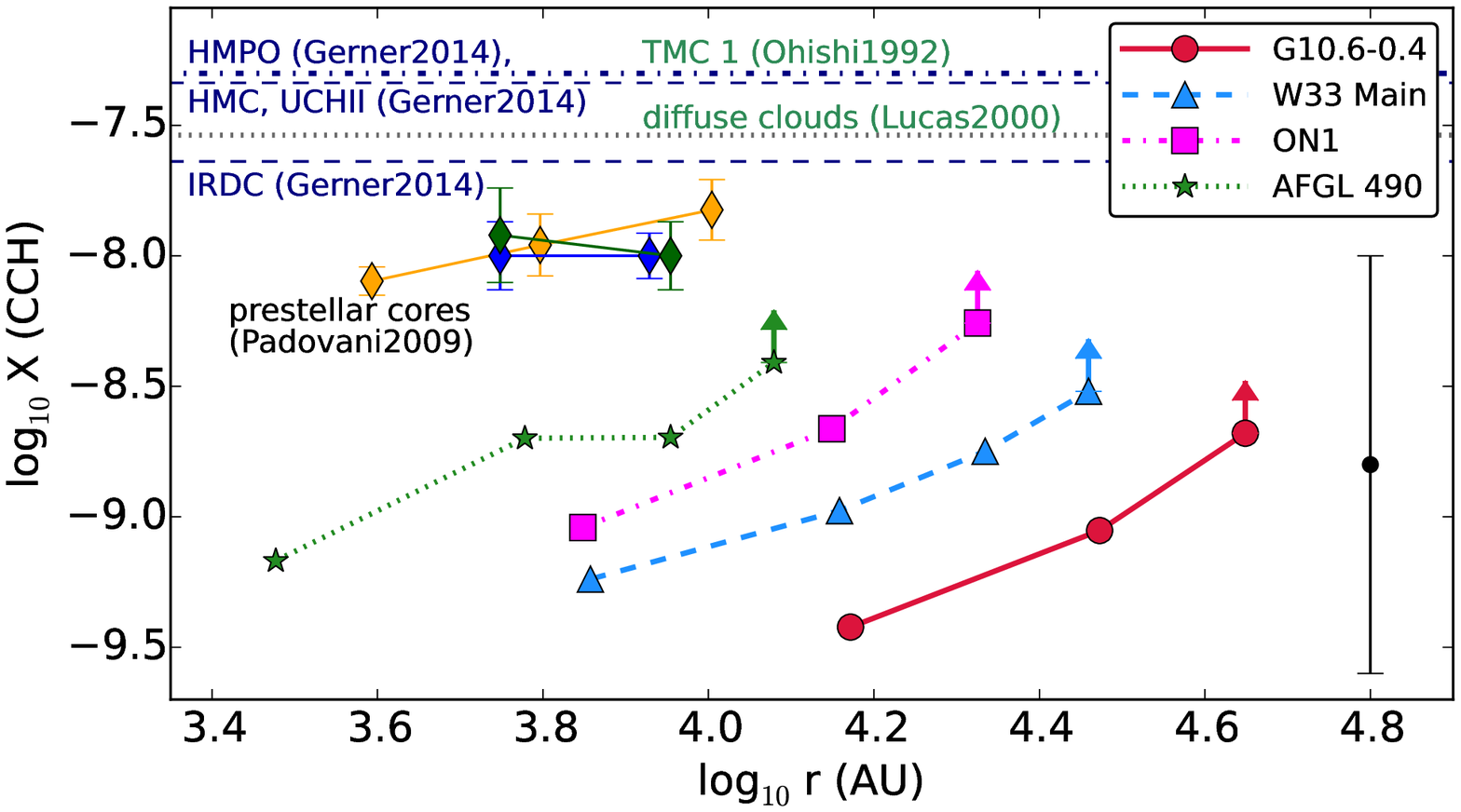} 
  \caption{ {\it Top:} C$_2$H column density radial profiles in logarithm scale of the four
  sources. Horizontal axis is the distance from the continuum peaks in units of
  AU. The typical error on the $N({\rm C_2H})$ is about 0.4 dex. The horizontal
  lines represent the $N({\rm C_2H})$ of other types of star forming
  regions \citep{Gerner:2014, Sakai:2010}. {\it Bottom:} C$_2$H abundance
  profiles. The points with arrows represent lower limits of the \cch
  abundances. The error bar in the bottom right corner denotes the typical
  error on the \cch abundances. On the top of the figure, the horizontal
  dashed lines represent the \cch abundances from other works
  \citep{Gerner:2014, Lee:1998, Lucas:2000}.  The diamonds on
  the upper left are the \cch abundance profiles of the prestellar cores
  \citep{Padovani:2009}.  \textit{Colorful figures are available in the online
  version.}}
  \label{fig:profiles2} 
\end{center} 
\end{figure}

Figure \ref{fig:profiles2} shows the $N({\rm C_2H})$ and $X({\rm C_2H})$ as a
function of the separation (physical scale) from the 1.1 mm continuum peaks.
For comparison, we also plot some $N({\rm C_2H})$ and $X({\rm C_2H})$ from
other works in Figure \ref{fig:profiles2}.  On the top panel, the horizontal
reference lines denote the median $N({\rm C_2H})$ of different types of star
formation regions obtained by \citet[][dashed]{Gerner:2014}, the $N({\rm
C_2H})$ of a low-mass star-forming region L1527 \citep[dotted,][]{Sakai:2010},
and TMC\,1 \citep[dot dashed,][]{Ohishi:1992}. Three of our sources have
similar C$_2$H column densities as the HMPO, but slightly lower than the
$N({\rm C_2H})$ of the UC\,H\textsc{ii} in \citet{Gerner:2014}. Note that
the reference lines trace very different spatial scales as they represent
single-dish observations. The $N({\rm C_2H})$ of AFGL\,490 is lower than other
sources, which might be due to the fact that the source is at a latter
evolutionary stage when many molecules are destroyed by the ionizing radiation.
For G10.6-0.4 and W33\,Main the $N({\rm C_2H})$ in the centers are about a factor of
two lower than their peak values, and overall the decreasing trend of $N({\rm
C_2H})$ in the core centers of our targets is not very obvious.

The bottom panel of Figure \ref{fig:profiles2} shows the $X({\rm C_2H})$ as a
function of radius. We found that the $X({\rm C_2H})$ at larger separation are
close to the values in the prestellar cores ($\sim$10$^{-8}$, denoted as
diamonds; \citealt{Padovani:2009}) but lower than those of other types of
clouds (reference lines). $X({\rm C_2H})$ systematically decreases toward the
1.1 mm continuum peaks by about one order of magnitude, and the four sources
have similar slopes of the C$_2$H abundance radial profiles. Comparing the
plots of $N({\rm C_2H})$ and $X({\rm C_2H})$ together we find that our derived
$X({\rm C_2H})$ are lower than that of other work, and the inconsistency
between our data and other work is mainly for the abundance plot.

\begin{figure}[t] 
  \begin{center}
  \includegraphics[angle=0,scale=0.43]{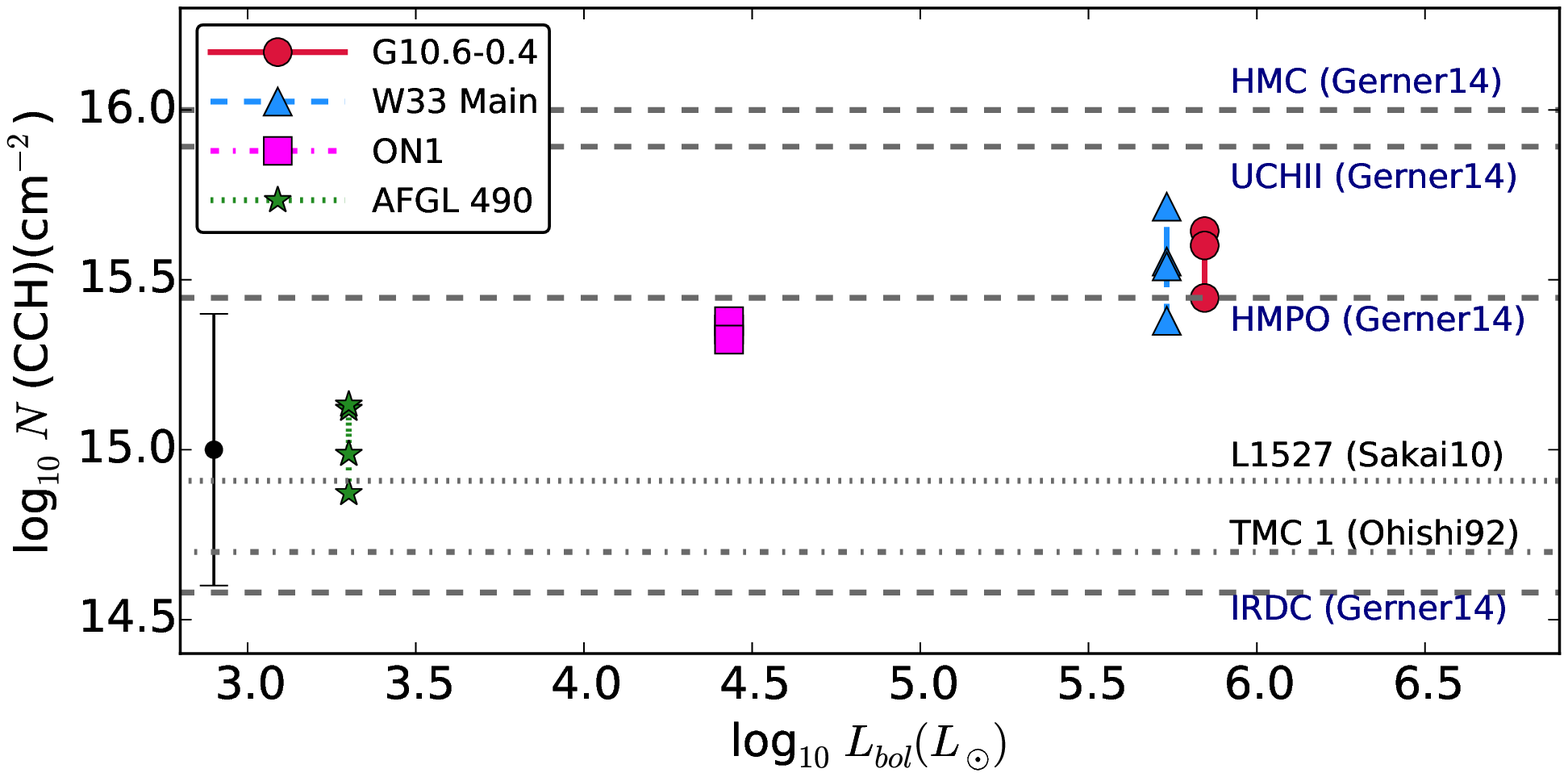} 
  \includegraphics[angle=0,scale=0.43]{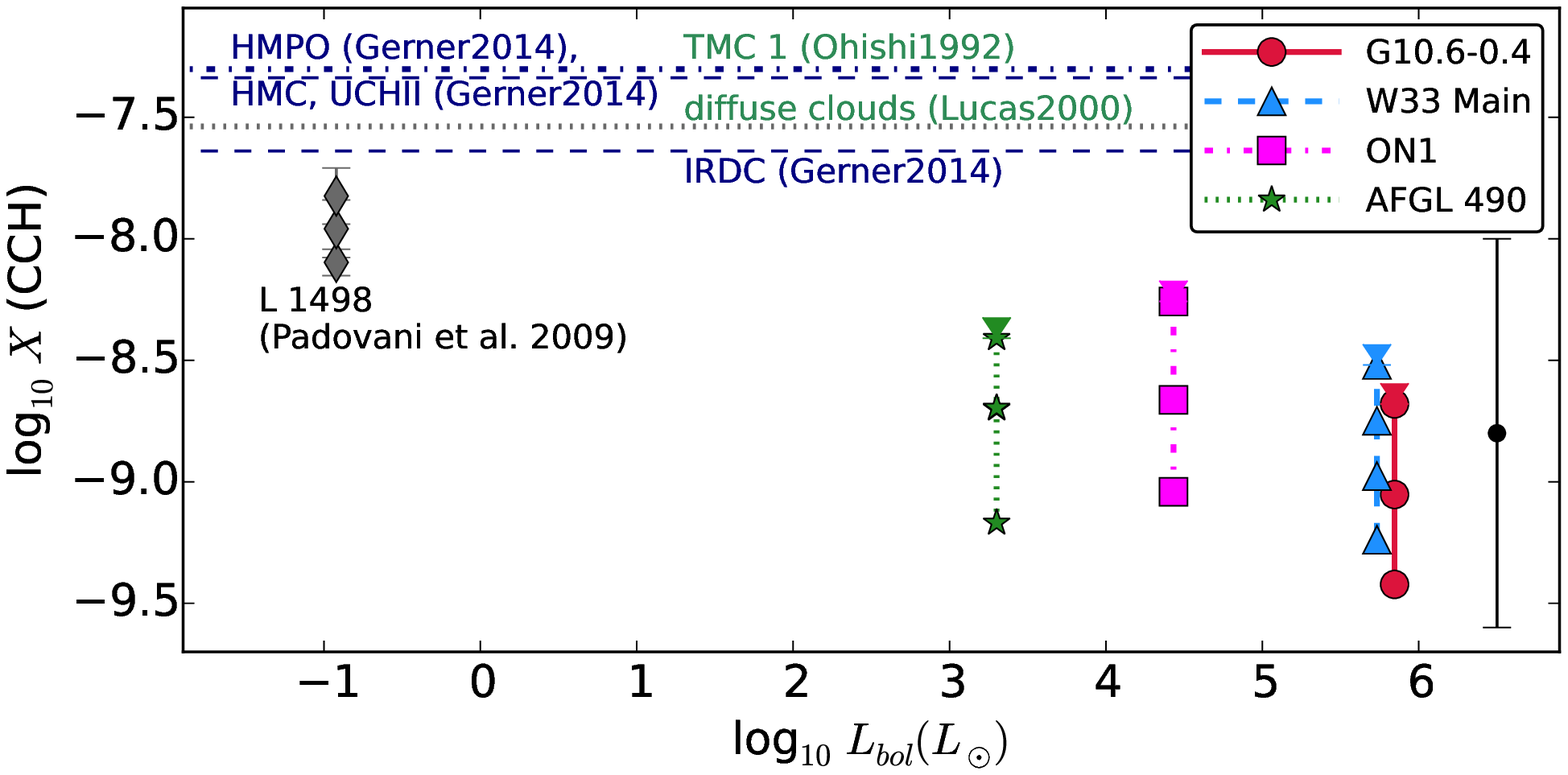} 
  \caption{{\it Top:} C$_2$H column density, $N({\rm C_2H})$, as a function of
  bolometric luminosity. The symbols and reference lines are the same as in Figure
  \ref{fig:profiles2}. The typical error on the $N({\rm C_2H})$ is about 0.4
  dex. {\it Bottom:} C$_2$H abundance $X({\rm C_2H})$ as a function of
  bolometric luminosity. The diamonds on the upper left are the \cch abundance
  of the prestellar cores L1498 \citep{Padovani:2009}, and its $L_{\rm bol}$ is 
  from \citet{Shirley:2005}.  
  \textit{Colorful figures are available in the online version.}}
  \label{fig:abundance_L} 
\end{center} 
\end{figure}

Figure \ref{fig:abundance_L} plots $N({\rm C_2H})$ and $X({\rm C_2H})$ as a
function of bolometric luminosities of the four sources. The top panel shows a
trend that $N({\rm C_2H})$ increases in more luminous targets, and the more
luminous sources G10.6-0.4 and W33\,Main have the largest $N({\rm C_2H})$,
despite the small sample and the uncertainty on $N({\rm C_2H})$. While the
luminosities of the sources differ by nearly three orders of magnitude, their
mean $N({\rm C_2H})$ differ only by about a factor of four. (see Section
\ref{sec:chemistry} for more discussion).  On the other hand, the bottom panel
of Figure \ref{fig:abundance_L} shows that $X({\rm C_2H})$ of our sample seem
to be lower than the less luminous source L1498, which is a prestellar core.
We again note that the uncertainty of $X({\rm C_2H})$ is large, although there
is a plausible trend in the bottom panel of Figure \ref{fig:profiles2} and
\ref{fig:abundance_L}.

%-------------------------  Discussion ---------------------

\section{Discussion and Conclusions} \label{sec:chemistry} 
We used the SMA to observe the C$_{2}$H $N$=3-2 emission lines ($\sim$262 GHz)
toward four high-mass star-forming regions G10.6-0.4, W33\,Main, ON\,1 and
AFGL\,490 with a spatial resolution better than 0.15 pc scale.  Our achieved resolution
is comparably smaller than the typical projected diameters of the HMCs, which
is the key to examining the C$_{2}$H chemical evolution under the
(proto)stellar illumination. Our results show that the extended C$_{2}$H
emission traces the dense gas envelopes in these high-mass sources, and
its column density $N({\rm C_2H})$ is of a few $10^{15}$cm$^{-2}$. 

Our observations show that, on $\sim 0.15$ pc scales, the C$_2$H emission peaks
are systematically offset from the high-mass star-forming hot cores. This is
consistent with the previous SMA observation of the source IRAS\,18089-1732
\citep{Beuther:2008}, as well as their model, which interprets such a scenario
as the transformation of C$_2$H into other molecules in the dense and hot
environment of the core center. The profiles in Figure \ref{fig:profiles1}
support such a proposed picture. Our findings of high column density C$_2$H are
also in agreement with the single-dish survey by \citet{Gerner:2014},  in more
luminous sources (Figure \ref{fig:abundance_L}). This might be a natural result
that more luminous sources have higher gas masses and a stronger PDR
(photo-dominated region) effect. The age of the specific HMPO model presented
in \citet{Beuther:2008} was 5$\times 10^4$ yr, but since the ages of our
observed sources are uncertain, and the abundance of C$_2$H cannot be derived
very accurately at this moment, a direct comparison between the models and the
observations remains difficult. 

In Figures \ref{fig:profiles2} and \ref{fig:abundance_L} we compare our derived
$N({\rm C_2H})$ with that from the previous observations toward a sample of
high-mass star-forming regions at different evolutionary stages \citep[HMPO,
HMC, and UC\,H\textsc{ii},][]{Gerner:2014}. All sources Except for AFGL\,490
show similar $N({\rm C_2H})$, which are close to the median $N({\rm C_2H})$ of
the HMPO, but slightly lower than that of the UC\,H{\sc ii}.  We note that in
the work by \citet{Gerner:2014} the C$_2$H data  of the HMCs and
UC\,H\textsc{ii} (which are treated as later stages) were not well explained by
the model, which might imply a more complicated chemistry of C$_{2}$H.  $X({\rm
C_2H})$ of the four sources are also compared with that of other clouds,
including that from \citet{Gerner:2014}, the dark clouds TMC-1 and L134N, the
prestellar cores, and the diffuse molecular clouds (Section
\ref{sec:abundance}).  The abundances derived for our targets are significantly
lower than those from other clouds, which is likely induced by the uncertainty
of the interferometric observation. But note that the $X({\rm C_2H})$ of the
prestellar cores \citep{Padovani:2009} are also lower than in those works, and
their resolutions are comparable to the SMA ($\sim$ 0.15 pc) as their targets
are much closer, so we suspect that the low $X({\rm C_2H})$ we find could also
be a result of the difference between the physical resolution of the works.

Figure \ref{fig:abundance_L} shows an increasing trend of $N({\rm C_2H})$ in
more luminous sources, while in AFGL\,490 the mean $N({\rm C_2H})$ is the lowest.
Although AFGL\,490 is still embedded in a molecular envelop, it is already in a
transition stage to Herbig Be stars \citep{Schreyer:2006}, so it is likely that
the neutral molecular gas content of AFGL\,490 has been reduced as a result of
the
ionized radiation.  In fact, the reason for this trend in Figure
\ref{fig:abundance_L} is likely that more luminous sources tend to have
more molecular gas, but as our images (Figure \ref{fig:mom0}) show, while
\cch is weaker in the central regions where it is possibly transformed, this
trend show that the overall \cch mass can maintain a certain amount in the gas
envelop surrounding the cores. This is also consistent with the picture that
C$_2$H is a good tracer of PDR, in the sense that it traces the cloud
surface illuminated by the stellar radiation \citep{Meier:2005}.  It
is also possible to compare this trend with that of other molecular lines to
analyze the corresponding chemical process or radiation effect on molecules.
Owing to the uncertainties mentioned above that
cannot be accurately accounted for at this moment, the quantitative analysis 
of $X({\rm C_2H})$ is less conclusive, and it is unclear yet whether the
abundance of \cch is significantly affected by the bolometric luminosity. 
Short-spacing data are needed to improve the accuracy of the abundance, and
higher-resolution observations can help to discern the genuine spatial
structure of C$_2$H in different environments, as well as to provide additional
information on the $\lesssim 10^4$\,AU scales where models of different
parameters start to deviate \citep{Beuther:2008}.

\subsection{Chemistry of hydrocarbons}
Observations toward the PDR region MonR2 found that C$_2$H is extended
compared to CH and c-C$_3$H$_2$, and the abundance of C$_2$H is constant over
a wide range of $G_0$ \citep[UV intensity; ][]{Pilleri:2013}. They demonstrated
that the behavior of C$_2$H is dominated by time-dependent effects and
gas-grain chemistry in low-density molecular envelopes. This scenario agrees
with our observations, which also show evidence that C$_2$H is stronger in the
molecular envelopes. 

The formation of C$_2$H is mainly determined by the formation and evolution of
ionized hydrocarbons, which involves the association of $\rm C^+$ and \Ht: $\rm
C^+ \rightarrow CH^+_n$ ($n=2,3,4,5$). CH$^+_n$ can be 
transformed into CH$_4$, then $\rm C_2H_2^+$ and $\rm C_2H_3^+$. C$_2$H 
is formed from the dissociated recombination of $\rm C_2H_2^+$ and $\rm
C_2H_3^+$ \citep[see, e.g.,][for details]{Turner:2000, Meier:2005, Beuther:2008}:
\begin{equation}\label{reaction1a}
\rm C_2H_2^+ + \textit{e}^- \rightarrow C_2H + H,
\end{equation}
\begin{equation}\label{reaction1b}
\rm C_2H_3^+ + \textit{e}^- \rightarrow C_2H + 2H/H_2,
\end{equation}

Other pathways of producing C$_2$H include the photodissociation of acetylene
(C$_2$H$_2$), and the reaction between CH$_2$ and carbon atoms
\citep{Sakai:2010}:
\begin{equation}\label{reaction2}
\rm C + CH_2 \rightarrow C_2H + H.
\end{equation}

On the other hand, C$_2$H is mainly destroyed by the reactions with
oxygen atoms:
\begin{equation}
\rm C_2H + O \rightarrow CO + CH.
\end{equation}
It can also be transformed into a heavier carbon chain such as C$_5$
\citep{Mookerjea:2012}.  Models addressing the variation of gas-phase [C]/[O]
abundance ratio \citep[e.g.,][]{Terzieva:1998} might help constrain the process
of production/depletion of C$_2$H in these sources.

It would be also interesting to compare C$_2$H with other simple carbon chain
molecules such as HC$_3$N, since they share the same chemical precursor $\rm
C_2H_2^+$ \citep{Wootten:1980}. Our observation covered HC$_3$N $J=30-29$
simultaneously but its high excitation energy ($E_u \sim$ 200\,K) makes it
difficult to compare its chemistry with C$_2$H $N=3-2$ ($E_u \sim$ 25 K)
because they trace the distinct physical conditions. Observations at 3\,mm band
would be able to cover both C$_2$H 1-0 (87.4 GHz) and \hcccn 10-9 (91 GHz, $E_u
\sim$ 24 K) simultaneously, and such comparison is more suitable for studying
the chemistry of carbon chain molecules in cold gas. The comparison between
C$_2$H and other molecular lines and related chemistry will be further
addressed in future work.

\subsection{Extragalactic perspective}
Our derived $N({\rm C_2H})$ can also be compared with
studies on external galaxies. Observation of 1\,kpc scale molecular complex in
M\,51 by \citet{Watanabe:2014} found that $N({\rm C_2H})$ is a few
times $10^{14}$ cm$^{-2}$ (emission size corrected, corresponding $X({\rm C_2H})$ is
about $\sim 4.5 - 5 \times10^{-9}$), which is about an order of magnitude
lower than the average $N({\rm C_2H})$ of the sources in our work. This might
suggest that while C$_2$H has higher column density near 
PDRs on $\sim$ pc scales, and observations on large scales ($\sim 1$ kpc)
would mainly collect ${\rm C_2H}$ emission from those diffuse molecular gas.

In the high-resolution line surveys by \citet{Meier:2005, Meier:2012,
Meier:2015}, C$_2$H appears to be bright (comparable to CS in NGC\,253) and
well traces the circumnuclear PDR regions of the nearby galaxies, and even
traces the molecular outflow in the nucleus of Maffei\,2 \citep{Meier:2012}.
The hyperfine structures of C$_2$H can help estimate the optical depth, and the
frequency of C$_2$H 1-0 is very close to HCN 1-0 and HCO$^+$ 1-0, so they can
be observed simultaneously, and their different critical densities and chemical
properties can be compared to diagnose the environment of galaxies.
\citet{Jiang:2011} have reported a sample of active galaxies detected in C$_2$H
with the IRAM 30m telescope, and with the aid of ALMA and other facilities, the
role of C$_2$H in studying extragalactic molecular gas will be further
explored. 

%-----------------------------------------------------------------------------------------------------------------------------------
\acknowledgements
%We thank the referee . . .
We acknowledge the SMA staff for their help during and after the observations.  
X.J. thanks Keping Qiu for informative discussions on the source ON\,1. 
This work is supported under the National Natural Science Foundation of China
(grants 11390373, 11273015, 11133001, and 11328301), the Strategic Priority Research
Program ``The Emergence of Cosmological Structures''
of the Chinese Academy of Sciences (grant
XDB09000000), the National Basic Research Program (973
program No. 2013CB834905), and Specialized  Research Fund for the Doctoral
Program of Higher Education (20100091110009). Z-Y.Z. acknowledges support from
the European Research Council (ERC) in the form of Advanced Grant {\sc
cosmicism}. This research made use of Matplotlib \citep{Hunter:2007} and APLpy,
an open-source plotting package for
Python hosted at http://aplpy.github.com.

%-----------------------------------------------------------------------------------------------------------------------------------
\end{CJK*}

\bibliography{bibtex_cch.bib}

%-----------------------------------------------------------------------------------------------------------------------------------
%  \newpage
%  \clearpage
%\begin{landscape}

%-----------------------------------------------------------------------------------------------------------------------------------

\end{document}